\documentclass[%
reprint,
superscriptaddress,
amsmath,amssymb,
aps,
]{revtex4-2}

\usepackage{graphicx}
\usepackage{dcolumn}
\usepackage{bm}
\usepackage{xr}

\begin{document}
	
	
	\title{Creating high-contrast patterns in multiple-scattering media via wavefront shaping}
	
	\author{Liam Shaughnessy$^{1\ast}$, Rohin E. McIntosh$^{1\ast}$, Arthur Goetschy$^{2}$, Chia~Wei~Hsu$^{3}$, Nicholas Bender$^{4}$, Hasan Y\i lmaz$^{5}$, Alexey Yamilov$^{6}$, and Hui Cao$^{\dagger}$ }
	
	\affiliation{Department of Applied Physics, Yale University, New Haven, Connecticut 06520, USA. \\  
		$^{2}$  Institut  Langevin, ESPCI  Paris, PSL  University,  CNRS, F-75005  Paris,  France \\
		$^{3}$ Ming Hsieh Department of Electrical and Computer Engineering, University of Southern California, Los Angeles, California 90089, USA \\
		$^{4}$  School of Applied and Engineering Physics, Cornell University, Ithaca, New York 14850, USA \\
		$^{5}$ Institute of Materials Science and Nanotechnology, National Nanotechnology Research Center (UNAM), Bilkent University, 06800 Ankara, Turkey \\
		$^{6}$ Physics Department, Missouri University of Science \& Technology, Rolla, Missouri \\
		$^\ast$ These two people contributed equally to this work. \\ $^\dagger$ hui.cao@yale.edu
	}

	\date{\today}
	
	\begin{abstract} 
		
		Wavefront shaping allows focusing light through or inside strongly scattering media, but the background intensity also increases due to long-range correlations, reducing the target's contrast. By manipulating non-local intensity correlations of scattered waves in a disordered system with input wavefront shaping, we create high-contrast patterns behind strongly scattering media and targeted energy delivery into a diffusive system with minimal change in the surrounding intensity. These are achieved by introducing the contrast operator and the difference operator, and utilizing their eigenstates to maximize the target-to-background intensity contrast and energy difference. This work opens the door to coherent control of non-local effects in wave transport for practical applications.

	\end{abstract}
	
	
	\maketitle
	
	

	Non-local correlations between multiply scattered waves are a hallmark of mesoscopic transport in diffusive systems \cite{sheng2007introduction, akkermans2007mesoscopic, carminati2021principles}. They exist in multiple domains, including space, time, frequency, angle, and polarization \cite{berkovits1994correlations, van1999multiple, dogariu2015electromagnetic}. For example, with varying incident wavefronts, the fluctuation of transmitted intensity at one location is correlated with that at another location of separation exceeding the average size of speckle grains (of the order of wavelength). Such long-range spatial correlations have direct consequences for controlling mesoscopic transport with wavefront shaping \cite{mosk2012controlling, rotter2017light, cao2022shaping}.
	On one hand, focusing light on a single speckle into or through a diffusive sample simultaneously increases the neighboring speckle intensity, hence limiting the target-to-background contrast \cite{vellekoop2008universal, davy2012focusing, cheng2014focusing}. 
	On the other hand, long-range correlations facilitate the total transmission enhancement and energy delivery to an extended target that contains many speckles with a single optimized incident wavefront \cite{kim2012maximal, popoff2014coherent, sarma2016control, hsu2015broadband, hsu2017correlation, bender2022depth}. 
	
	For many applications including optical communication, photothermal therapy, microsurgery, and optogenetics~\cite{2004_Yanik, 2011_Fenno, 2015_Park_R, 2017_Ruan_SA, 2017_Pegard_NC}, it is important not only to maximize light intensity over an extended target but also to optimize the target-to-background intensity contrast. While the former benefits from positive intensity correlations within the target region, the latter requires suppression of positive correlations or even introduction of negative correlation between intensity inside and outside the target region. Whether this is possible to achieve, given the constraints from non-local correlations, is unclear, as it requires manipulation of non-local correlations. Previously, long-range intensity correlations inside diffusive waveguides were modified by tailoring the waveguide geometry~\cite{sarma2015using}. However, it is not always possible to reconfigure the boundary shape of a scattering sample in practical applications. Here we explore a different approach: we shape the input wavefront to manipulate the non-local correlations of scattered waves.
	
	In this work, we introduce two operators: the contrast operator and the difference operator. We demonstrate that the contrast operator's eigenvectors can maximize the target-to-background intensity contrast. Strong intensity attenuation in a chosen surrounding area enables the generation of high-contrast patterns behind the scattering medium by shaping the incident wavefront of a coherent beam (Fig.~1). However, the slight intensity increase on the target may not be optimal for applications. To enhance the target intensity as much as possible with little change in the surrounding intensity, we introduce the difference operator. Its eigenvectors are used for broad-area focusing behind a thick scattering sample and for targeted energy delivery deep inside a diffusive waveguide. Using perturbation theory, we provide a physical explanation of the experimental data. 
	
	\begin{figure}[ht]
		\centering
		\includegraphics[width=.48\textwidth]{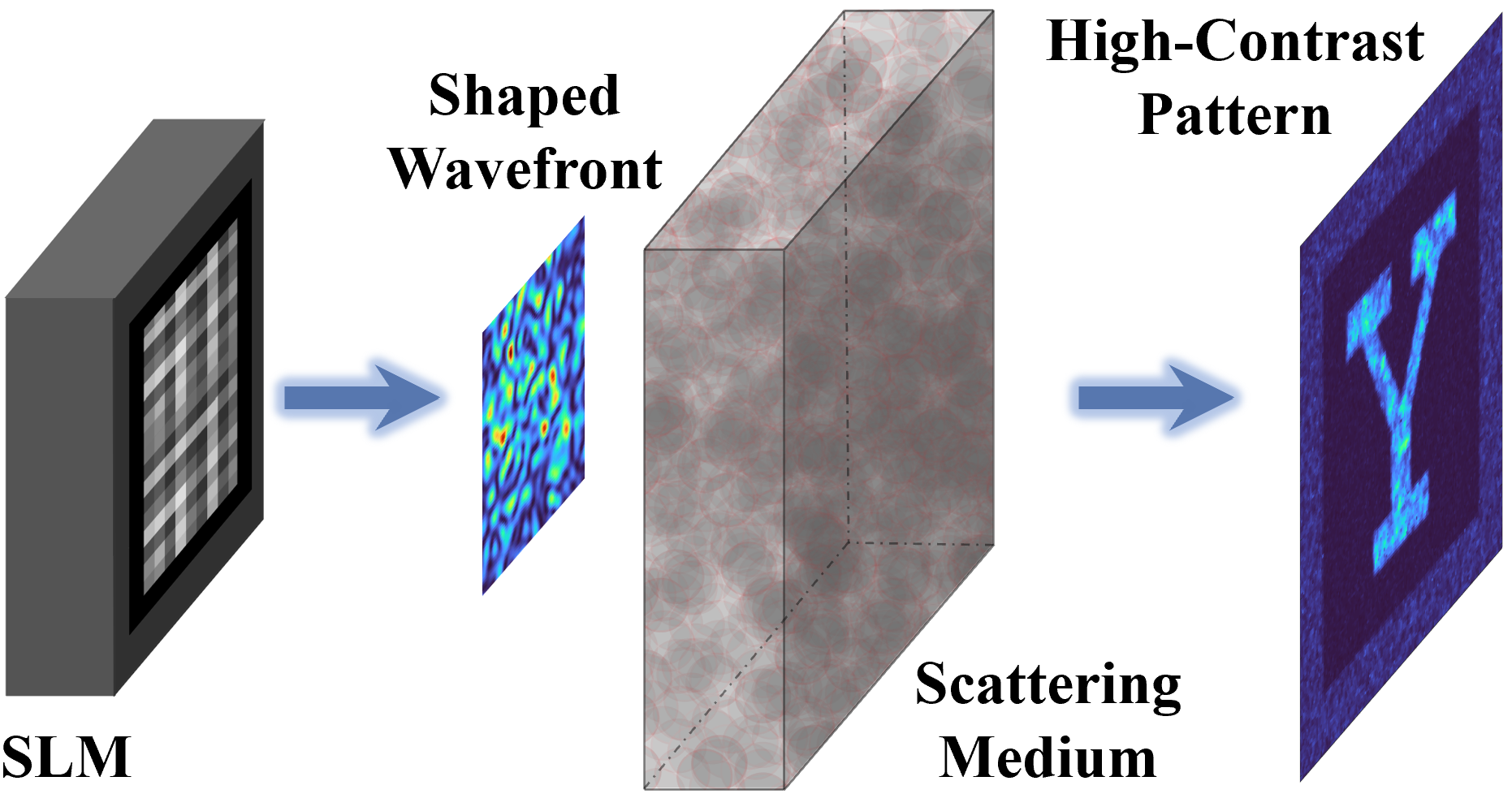}
		\caption{\label{volume} {\bf Schematic for the propagation of an eigenstate of the contrast operator}. A high-contrast pattern with a sharp boundary is created behind a disordered slab by shaping the incident wavefront of a laser beam with a spatial light modulator (SLM). }
	\end{figure}    
	
	We first investigate the focusing of light onto an extended area behind a three-dimensional (3D) scattering slab. The sample consists of ZnO nanoparticles that are spin-coated on a cover slide. The ZnO layer thickness is about 60 $\mu$m, much larger than the transport mean free path $\sim 1$ $\mu$m~\cite{hsu2017correlation}. The average light transmittance at the probe wavelength of $\lambda$ = 532 nm is approximately 0.03. As schematically shown in Fig.~1, the incident wavefront of a laser beam (continuous-wave at $\lambda$ = 532 nm) is modulated by a phase-only spatial light modulator (SLM) before illuminating the ZnO slab. The illumination spot on the sample surface has a diameter of 24 $\mu$m. The number of input spatial channels modulated by the SLM is 2048 for two orthogonal polarizations. A CCD camera is placed at the Fourier plane of the output surface of the slab to measure the transmitted intensity patterns. To eliminate short-range correlations among output channels, the transmitted fields are sampled at one point per speckle grain. Using a phase-stepping common-path interferometric method, we measure the (partial) transmission matrix~\cite{hsu2017correlation}.
	
	\begin{figure*}[ht]
		\centering
		\includegraphics[width=.98\textwidth]{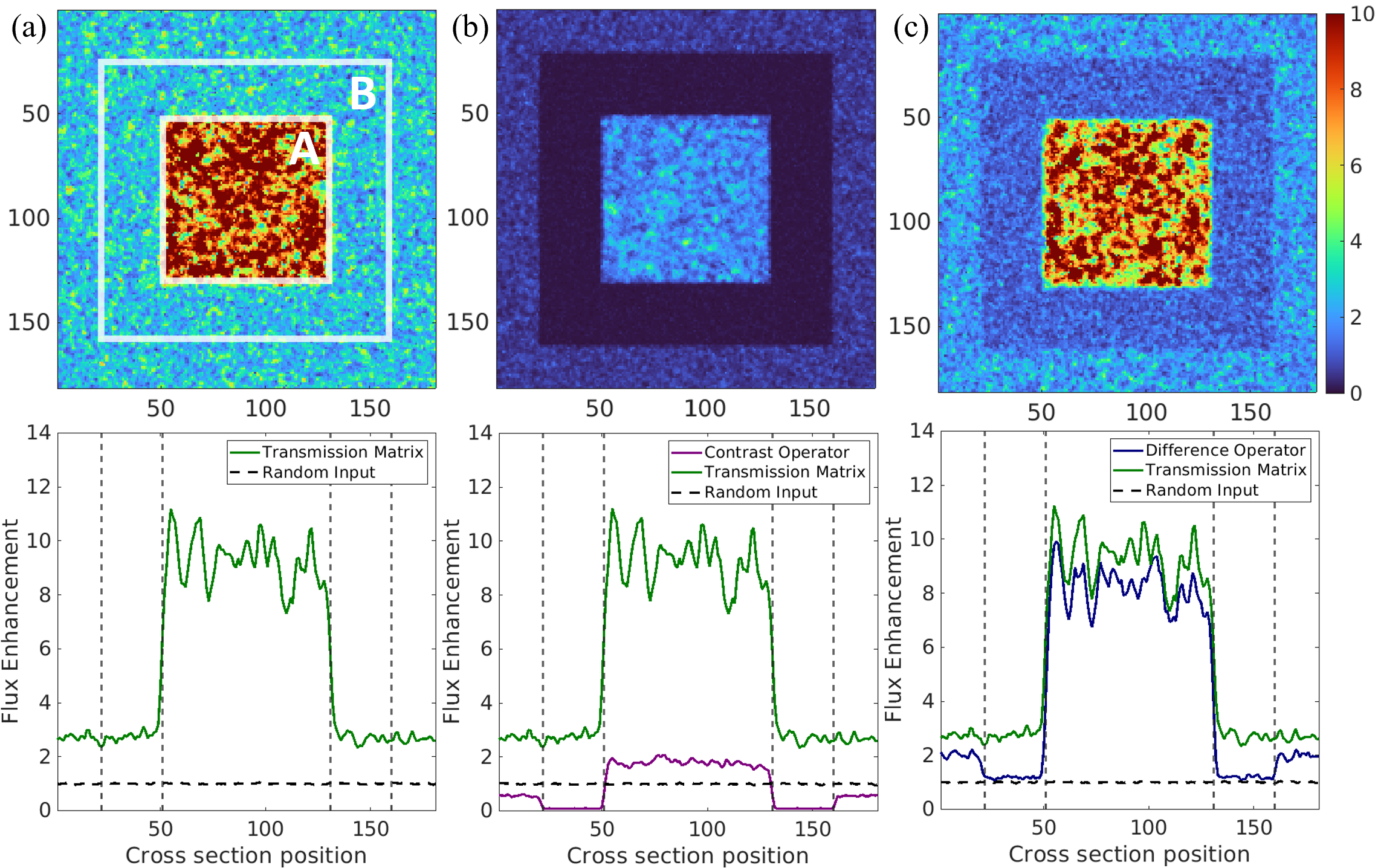}
		\caption{\label{volume} {\bf Eigenstates of transmission, contrast, and difference operators}. Intensity distributions of transmitted light through a multiple-scattering slab for eigenstates with maximal flux inside the target region A (a), highest contrast of flux in A to that in a surrounding region B (b), largest flux difference between A and B (c). Lower panels show intensity across a cross-section of the images in the upper panels. The green curve represents the state in (a), purple in (b), blue in (c), and black dashed line for random input wavefronts. The square A contains $\sim876$ speckle grains, and ring B contains $\sim1752$ speckles.}
	\end{figure*}   
	
	In Fig.~2a, the target region (labeled A) is a square at the center of the field of view containing 876 speckle grains. The matrix $t_A$ maps the incident fields to transmitted fields inside area A. The eigenvector of $t_A^\dagger t_A$ with the largest eigenvalue gives the optimal incident wavefront for focusing transmitted light to target A \cite{hsu2017correlation}. While the mean intensity within A is almost 10 times that under random wavefront illumination, the intensity outside A is more than doubled due to long-range correlations. 
	
	To suppress the intensity increase in a surrounding area (labeled B in Fig.~2a), we introduce the contrast operator $(t_B^\dagger t_B)^{-1} \,  t_A^\dagger t_A $, where $t_B$ denotes the matrix that maps incident fields to transmitted fields in B. Figure 2 shows an example with the area ratio of B to A equal to 2. From the definition of the contrast operator, its largest eigenvalue is equal to the maximal ratio of flux (spatially integrated intensity) in A to B, and the corresponding eigenvector gives the input wavefront. As shown in Fig.~2b, the intensity in B nearly vanishes, while the intensity in A is slightly above the value for random wavefront illumination. Such changes are caused by strong destructive interference of scattered waves in B and weak constructive interference in A. 
	There is a sharp transition between A and B: the intensity changes abruptly across the boundary of A and B. The opposite changes of intensity in A and B are surprising, because of the widely-known positive intensity correlations of multiply scattered light. This is a result of the wave interference effect. Hence, the spatial intensity correlations can be overcome by manipulating interference effects with wavefront shaping. 
	
	The maximal target-to-background contrast in Fig.~2b is achieved mostly by suppressing the background intensity instead of enhancing the target intensity. Some applications like photodynamic therapy and microsurgery benefit more from intensity enhancement of the target than suppression of background intensity. To keep the background intensity nearly unchanged from that with random wavefront illumination while increasing the target intensity as much as possible, we introduce the difference operator $  t_A^\dagger t_A  - t_B^\dagger t_B$. Its maximal eigenvalue is the largest possible flux difference between area A and B and the corresponding eigenvector gives the incident wavefront. Figure 2c shows the target intensity is reduced slightly from focusing to A, meanwhile, the surrounding intensity barely deviates from that with random wavefront illumination.
	
	\begin{figure}[ht]
		\centering
		\includegraphics[width=.48\textwidth]{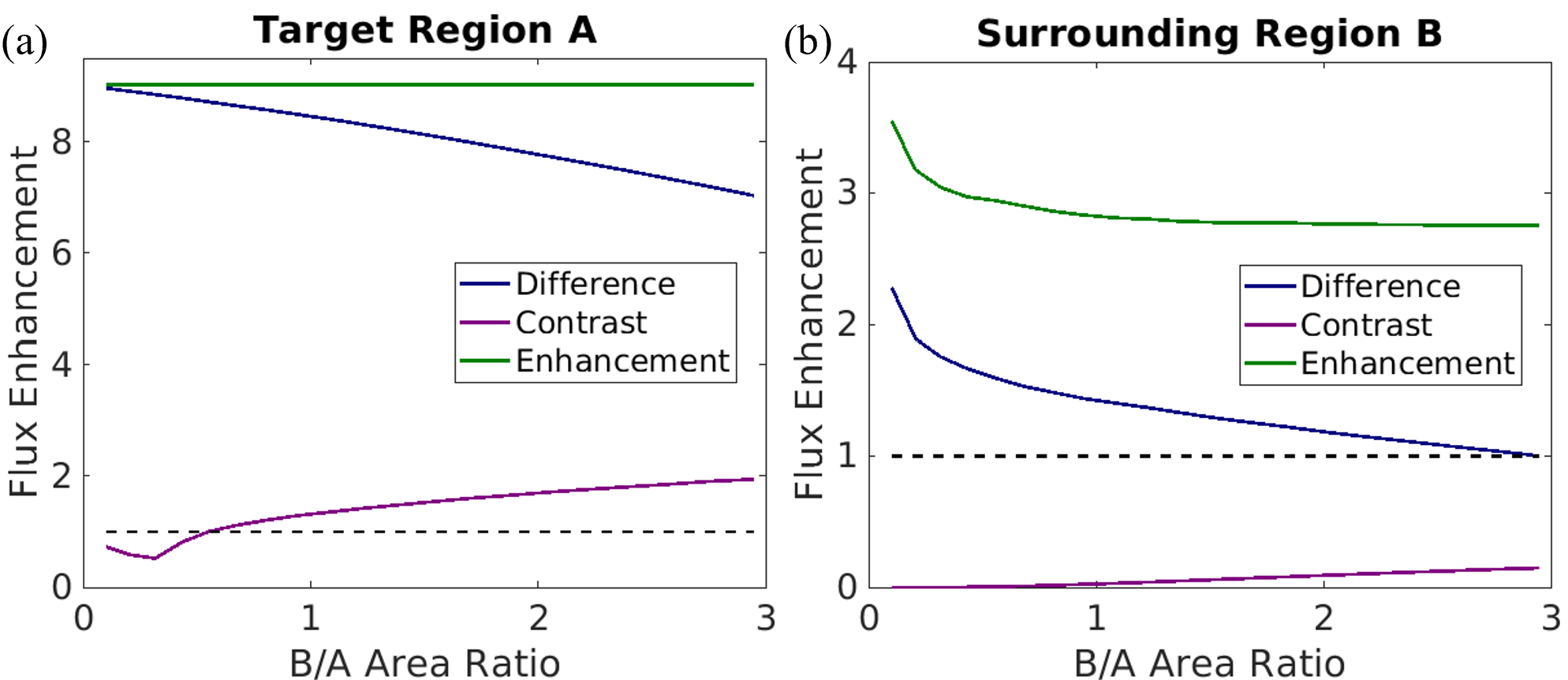}
		\caption{\label{volume} {\bf Comparison of contrast and difference eigenstates}. Fluxes in target A (a) and surrounding region B (b) as a function of the area ratio of B over A. The contrast eigenstate (magenta) has a flux increase in A and B with B/A ratio, while the difference eigenstate (blue) has a flux decrease in both regions. The black dashed line corresponds to random input wavefronts.}
	\end{figure}
	
	With the target region A being fixed, we vary the surrounding area B and compare the contrast eigenstates to the difference eigenstates. As the area ratio of B over A increases, the highest-contrast state features intensity growth in both A and B (Fig.~3). The attenuation of surrounding intensity becomes less efficient for larger B, leading to lower target-to-background contrast. The largest-difference state exhibits opposite behavior: intensities in A and B both drop with increasing B/A ratio (Fig.~3). At a small B/A ratio, the flux difference between A and B is maximized mostly by enhancing the target intensity, because A dominates over B. For a large B/A ratio, the maximal flux difference is achieved by avoiding the increase of background intensity, as B becomes dominant over A. 
	
	In addition to controlling transmission, targeted energy delivery deep into a multiple-scattering system has important applications. We show next that the difference operator is also useful for delivering energy to an extended target deep inside a two-dimensional (2D) diffusive waveguide by incident wavefront shaping. The planar waveguide structure is fabricated in a silicon-on-insulator wafer by electron beam lithography and reactive plasma etching \cite{sarma2016control}. As shown in Fig.~4a, 100 nm diameter holes are randomly distributed in the waveguide, which has photonic crystal sidewalls to reflect light. The waveguide is 15 \textmu m wide, supporting 55 propagating modes at the probe wavelength $\lambda = 1.55$~\textmu m. The transport mean free path $\ell_t = 3.2$~\textmu m is much shorter than the disordered region length $L = 50$~\textmu m in the waveguide \cite{yamilov2014position}. Hence, incident light from one end of the waveguide undergoes multiple scattering and diffusive transport through the waveguide. A small amount of light scatters out-of-plane from the air holes, providing a direct probe of field distribution inside the disordered region. While the material absorption is negligible, the out-of-plane scattering can be modeled as an effective loss and accounted for in the diffusive dissipation length $\xi_a$ = 28~\textmu m \cite{sarma2016control}. 
	
	Using an SLM, we shape the one-dimensional (1D) incident wavefront of a continuous-wave laser beam at $\lambda = 1.55$~\textmu m before launching the light through the edge of the wafer into a ridge waveguide that is connected to the disordered waveguide. The out-of-plane scattered light overlaps with a reference beam of a flat phase front, and a camera records their interference pattern. From the image, the internal field distribution is reconstructed with a spatial resolution of 1.1~$\mu$m. 
	
	
	\begin{figure}[ht]
		\centering
		\includegraphics[width=.48\textwidth]{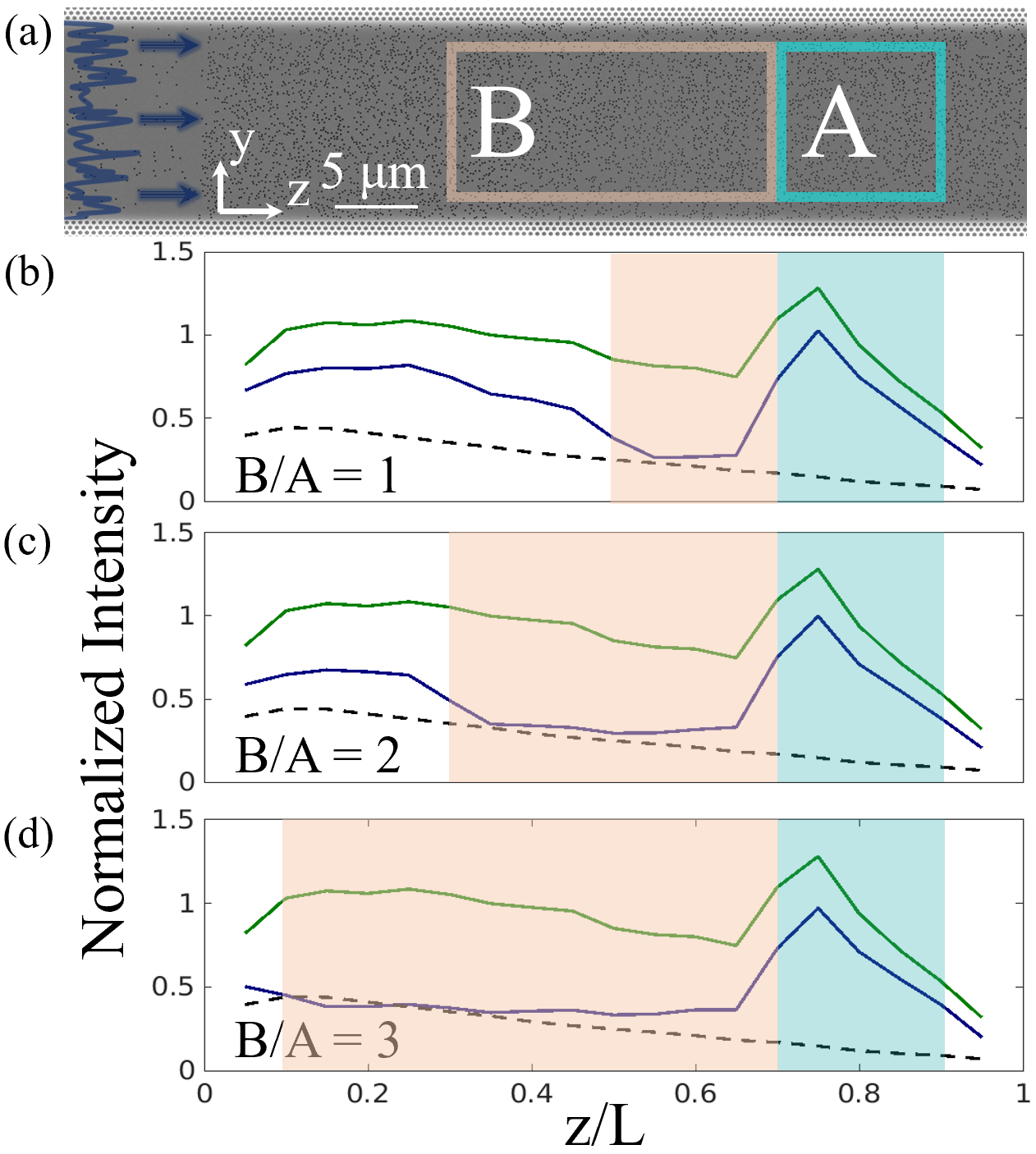}
		\caption{\label{volume} {\bf Difference eigenstate for targeted energy delivery into a diffusive waveguide}. (a) Shaping the wavefront of incident light into a disordered waveguide enhances the energy within an extended target A (shaded yellow) while minimizing the intensity increase in region B (shaded cyan) right in front of A. The silicon waveguide has a random array of air holes over a region of length 50 $\mu$m and width 15 $\mu$m.  Target A is a $10 \times 10$ $\mu$m$^2$ square centered at depth 40 \textmu m. Region B is 10 $\mu$m wide, and its length varies from 10 $\mu$m to 30 $\mu$m. (b-d) Cross-section averaged intensities for B/A area ratio equal to 1 (b), 2 (c), 3 (d).  The deposition eigenstates (green) increase intensities in both A and B, while the difference eigenstates (blue) enhance intensity in A with little change of intensity in B from random input wavefronts (black dashed).}
	\end{figure}
	
	In Fig.~4, we choose the target region A to be a 10~\textmu m $\times$ 10 \textmu m square centered at depth $z_D$ = 40~\textmu m.  From the interferometric measurement, we retrieve the scattered field everywhere inside the disordered waveguide and construct the deposition matrix $\mathcal{Z}_A$ that maps the incident fields to the fields within the target region \cite{bender2022depth}. The largest eigenvalue of $\mathcal{Z}_A^\dagger \mathcal{Z}_A$ gives the maximal possible energy (spatially integrated intensity) that can be delivered to the target area A. However, the intensity outside A is notably higher than that under random wavefront illumination, as shown by the green curve in Fig.~4. On one hand, long-range correlation allows a single input wavefront to enhance intensity everywhere within A. On the other hand, it also causes an intensity increase outside A. 
	
	To suppress the intensity enhancement right before A, we select region B as a 10 \textmu m $\times$ 10 \textmu m square centered at depth $z_D$ = 30 \textmu m, as illustrated in Fig.~4a. The deposition matrix to region B is $\mathcal{Z}_B$. To prevent the intensity in B from increasing with that in A, we use the difference operator $\mathcal{Z}_A^\dagger \mathcal{Z}_A - \mathcal{Z}_B^\dagger \mathcal{Z}_B$. Its eigenstate with maximal eigenvalue keeps the intensity in B almost equal to the value of random inputs while still dramatically raising the intensity in A (blue curve in Fig.~4b). We further increase the length of region B and modify the difference operator $\mathcal{D} = \mathcal{Z}_A^\dagger \mathcal{Z}_A - \gamma \mathcal{Z}_B^\dagger \mathcal{Z}_B$, where $\gamma$ is the area ratio of A to B. Incorporating $\gamma$ to the difference operator would reduce the contribution from B when its area exceeds A so that the maximal difference between A and B is achieved by enhancing energy in A ($U_A$) more than by suppressing energy in B ($U_B$). In Figs.~4c and ~4d, the maximal difference eigenstates, for area B equal to twice and even three times area A, feature significant intensity enhancement in A while the intensity in B barely increases. 
	
	Comparing the maximal difference eigenstate to the maximal deposition eigenstates, we find that the energy reduction in B ($U_B$) is notably larger than that in A ($U_A$). This can be qualitatively understood as follows. The largest eigenvalue of $\mathcal{D}$ maximizes $U_A - \gamma U_B$, while the maximal eigenvalue of $\mathcal{Z}_A^\dagger \mathcal{Z}_A$ gives the highest possible $U_A$. Since it is impossible to increase $U_A$ any further from the maximal deposition eigenvalue, the only way to enhance the difference between $U_A$ and $U_B$ is to reduce $U_B$ much more than $U_A$ for the maximal difference state. Thus the background intensity increase, seen in the maximal deposition eigenstate, is suppressed in the maximal difference state. 
	
	\begin{figure}[ht]
		\centering
		\includegraphics[width=.48\textwidth]{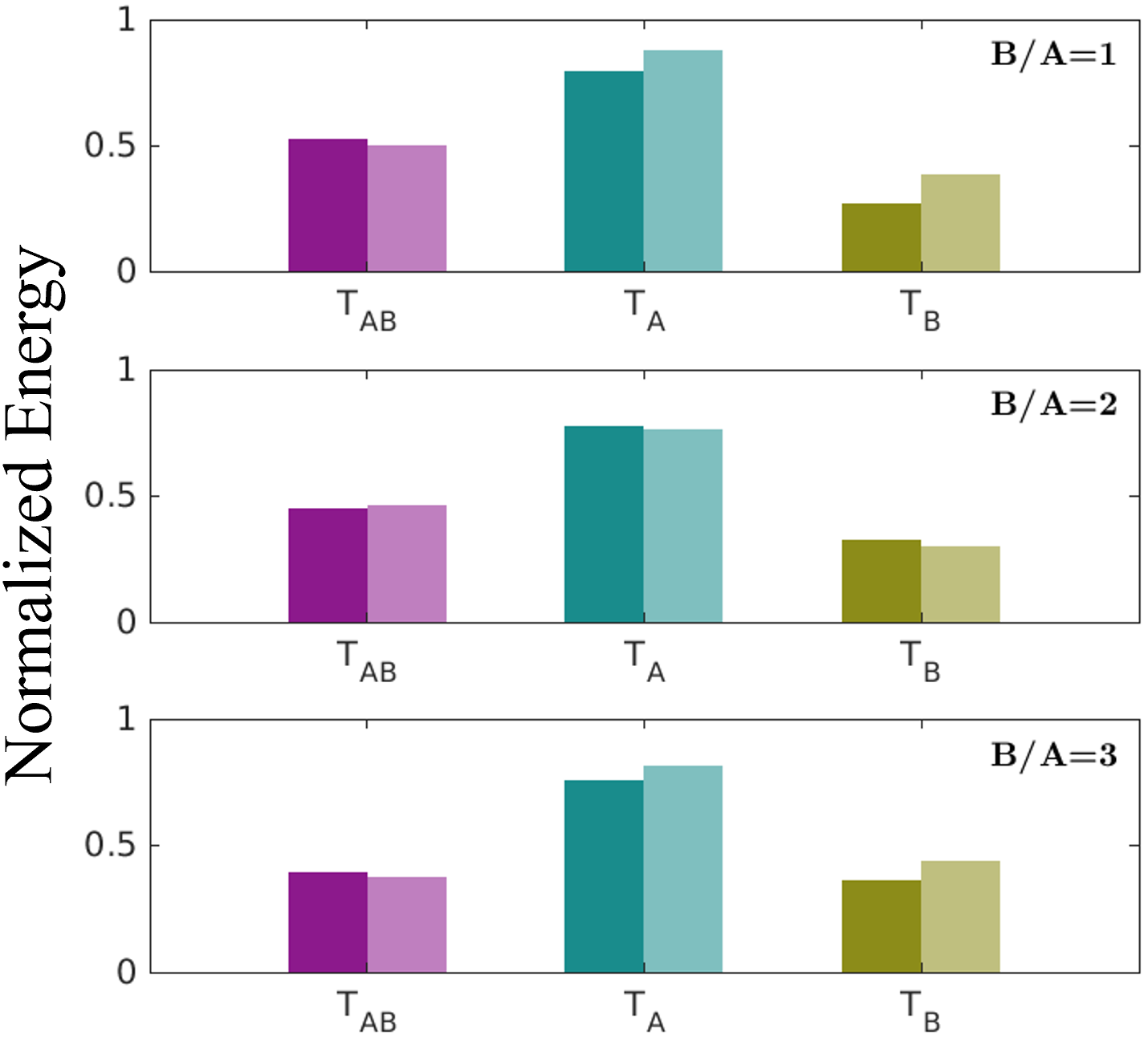}
		\caption{\label{volume} {\bf Perturbative approach to the difference operator}. The largest eigenvalue of the difference operator (left column), energy in target A (middle column), and energy in region B (right column), calculated with perturbation theory (light color), are close to the actual values (dark color). The area ratio of B to A is 1 in (a), 2 in (b), and 3 in (c). }
	\end{figure}
	
	To be more quantitative, we develop a perturbation model by treating $\gamma \mathcal{Z}_B^\dagger \mathcal{Z}_B$ as a perturbation to $\mathcal{Z}_A^\dagger \mathcal{Z}_A$. This approach is valid for the maximal difference eigenstate, which typically satisfies $U_A \gg \gamma U_B$. To have enough accuracy, we perform a third-order perturbation expansion to calculate the maximal difference eigenvalue, as well as the energy deposited in the areas A and B by the corresponding eigenstate~\cite{SI}. The left column in Fig.~5 shows that the maximal eigenvalue obtained by the perturbation calculation almost equals the actual value.
	In addition, the calculated energies in A and B~\cite{SI} 
	are close to the experimental values for varying B/A ratios, as seen in the middle and right-hand columns of Fig.~5. Our model predicts that the maximum difference eigenstate exhibits a $U_B$ reduction relative to the maximum deposition eigenstate greater than the $U_A$ reduction by a factor of the order of $2/\gamma$, consistent with our data. The relative change of intensity inside B can be expressed as~\cite{SI}
	\begin{equation}
	\frac{\Delta I_B}{I_B} \ge \left( \frac{2}{\gamma} \frac{N_A}{N_B} \frac{I_A}{I_B} \right ) \frac{\Delta I_A}{I_A},
	\end{equation}
	where $N_A$ and $N_B$ are the numbers of speckle grains inside the areas A and B. This shows that the relative change of intensity in B is always much larger than the relative variation of intensity in A, in the regime where the perturbation approach holds ($\gamma N_B \le N_A$). These predictions quantitatively support the observations shown in Fig.~4.

	In summary, by shaping the incident wavefront of a coherent beam, we manipulate nonlocal intensity correlations of multiply-scattered waves in 2D diffusive waveguides and 3D scattering slabs. Experimentally we demonstrate simultaneous enhancement of the energy delivered to an extended target and the target-to-background contrast.  By introducing the contrast operator and the difference operator, we find the maximal contrast and difference of energies between the target region and a surrounding area. While the maximal contrast state strongly attenuates the background intensity, the maximal difference state enhances the target intensity while avoiding the background increase. These results open the door to controlling nonlocal effects in the mesoscopic transport of waves for practical applications based on targeted energy delivery behind or inside strong-scattering media.

	This work is supported partly by the National Science Foundation (NSF) under Grant No. DMR-1905465 and by the Office of Naval Research (ONR) under Grant No. N00014-221-1-2026.
	
	\nocite{*}
	
	\bibliography{main}
	
\end{document}